\documentclass[apjl]{emulateapj}
\usepackage{apjfonts}

\begin{document}
\journalinfo{Astrophysical Journal Letters}
\shortauthors{Fish et al.}
\shorttitle{Excited OH in Supernova Remnants and W3(OH)}
\title{Effelsberg Observations of Excited-State (6.0 GH\lowercase{z})
OH in Supernova Remnants and W3(OH)}
\author{ Vincent L.\ Fish\altaffilmark{1,2},
         Lor\'{a}nt O.\ Sjouwerman\altaffilmark{1},
      \& Ylva M.\ Pihlstr\"{o}m\altaffilmark{3}}
\altaffiltext{1}{National Radio Astronomy Observatory, 1003 Lopezville
Rd., Socorro, NM 87801}
\altaffiltext{2}{Jansky Fellow}
\altaffiltext{3}{Department of Physics and Astronomy, University of
  New Mexico, Albuquerque, NM 87131}
\begin{abstract}
While masers in the 1720~MHz transition of OH are detected toward many
supernova remnants (SNRs), no other OH transition is seen as a maser
in SNRs.
We present a search for masers at 6049~MHz, which has recently been
predicted to produce masers by pure collisional excitation at
conditions similar to that required for 1720~MHz masing.
The Effelsberg 100 m telescope was used to observe the excited-state
6016, 6030, 6035, and 6049~MHz lines of OH toward selected SNRs, most
of which have previously-detected bright 1720~MHz masers.
No excited-state masers are found toward SNRs, consistent with
previous observations of the 6049~MHz and other excited-state
transitions.  We do not see clear evidence of absorption toward SNR
target positions, although we do see evidence of absorption in the
molecular cloud at $+50$~km\,s$^{-1}$ near Sgr~A East.  Weak
absorption is detected at 6016~MHz toward W3(OH), while stronger,
narrower emission is seen at 6049~MHz, suggesting that the 6049~MHz
emission is a low-gain maser.
We conclude that conditions in SNRs are not conducive to excited-state
maser emission, especially in excited-state satellite lines.
\end{abstract}
\keywords{masers --- supernova remnants --- ISM: individual (W3(OH),
  Sgr~A East) --- radio lines: ISM --- radiation mechanisms:
  non-thermal}

\section{Introduction}

About two dozen supernova remnants (SNRs) are known to host 1720~MHz
OH masers \citep[][and references therein]{green06}.  The excitation
mechanism for 1720~MHz masers in SNRs is commonly believed to be
collisional excitation from a C-type shock \citep{lockett99,wardle99}.
Other ground-state OH transitions are sometimes seen in absorption
\citep[e.g.,][]{goss68}, which can be helpful for modelling the
physical conditions in the 1720~MHz masing region \citep{hewitt06}.
However, to date no other OH transition, nor a transition of any other
molecule, has been detected as a maser associated with an SNR.  The
detection of a second maser transition in SNRs would place strong
constraints on physical conditions (density, temperature, OH fraction,
ortho-to-para H$_2$ ratio, etc.) in the masing region, especially if
spatial coincidence with a 1720~MHz maser was observed.  Excited-state
transitions, which occur at higher frequencies, would be especially
useful if detected toward the Galactic center, where angular
scattering is large \citep[see \S~1 of][]{pihlstrom07}.

Another motivation for finding a second masing transition in SNRs is
to confirm the Zeeman interpretation of splitting seen between left
and right circular (LCP, RCP) components at 1720~MHz.  Differences in
LCP and RCP velocities are used to compute magnetic field strengths in
SNRs \citep[e.g.,][]{brogan00}, important for quantifying magnetic
pressures.  However, \citet{elitzur96,elitzur98} shows that circular
polarization can be generated by non-Zeeman mechanisms when the maser
is saturated and splitting is small compared to a line width, as is
the case in SNR 1720~MHz masers.  While there are theoretical reasons
to believe the Zeeman interpretation of SNR 1720~MHz maser splitting
\citep[see \S~4.1 of][]{brogan00}, direct confirmation of the magnetic
field strengths from a second OH transition would lend greater
confidence in the derived magnetic field strengths.

Theoretical modelling predicts that collisions alone can excite
several transitions of OH.  In the low-density ($n \sim
10^5$~cm$^{-3}$) regime, only the 1720~MHz transition is inverted, but
at higher number and/or column densities the 6049~MHz transition also
becomes inverted, with some overlap in parameter space allowing
simultaneous inversion of the 1720 and 6049~MHz lines
\citep{pavlakis00,wardle07,pihlstrom07}.  At still higher densities
($n \sim 5~\times~10^6$~cm$^{-3}$), the 4765 and 1612~MHz transitions
become inverted \citep{pavlakis96,pihlstrom07}.  The 6035~MHz line may
also become inverted at high temperatures \citep[$T \sim
200$~K;][]{pavlakis00}.  Regardless of the details of the pump
mechanism, weak inversion (emission) is predicted in the 6049~MHz and
weak anti-inversion (enhanced absorption) in the 6016~MHz line from
the structure of the OH level diagram alone when infrared trapping
becomes important \citep{litvak69,elitzur77}.

Few observations have sought excited-state OH masers in SNRs.
\citet{mcdonnell07} looked for 6030, 6035, and 6049~MHz OH maser
emission toward southern SNRs but did not detect any 6049~MHz
emission.  Although they did find three new 6030 and 6035~MHz maser
sources, \citet{mcdonnell07} believe that these masers are associated
with \ion{H}{2} regions and therefore represent emission from
star-forming regions (SFRs), not SNRs.  In states of higher
excitation, \citet{pihlstrom07} searched for emission in the 4.7~GHz
$\Lambda$-doubling triplet and the 7.8 and 8.2~GHz quadruplets toward
four SNR complexes (as well as the 23.8~GHz quadruplet toward
Sgr~A~East) but did not detect any emission.  \citet{pihlstrom07} also
cross-correlated positions of single-peaked and irregular-spectrum
1612~MHz masers from blind surveys in the literature with positions of
known SNRs but found no probable associations.

For small velocity differences along the amplification path, the
6049~MHz transition is theoretically the most promising transition in
which to find new OH SNR masers, given that the collisional excitation
conditions require only slightly higher densities than those that
produce 1720~MHz masers.  We report on targeted observations of the
four 6.0~GHz lines mostly toward SNRs with bright 1720~MHz masers.

\section{Observations}

We observed 32 target positions with the Effelsberg 100 m telescope in
two runs on 2007~July~12 and 13.  Pointing centers are given in
Table~\ref{source-table}.  Observations were taken in
position-switched mode using the 5 cm primary focus receiver.  The
full width at half maximum (FWHM) beam size of the telescope is
approximately 130\arcsec\ at the 6.0~GHz frequencies of OH.  The
assumed sensitivity of the telescope is 1.55~K\,Jy$^{-1}$.

Data were collected using both the AK90 8192-channel autocorrelator as
well as the Fast Fourier Transform Spectrometer (FFTS).  The AK90 was
configured to produce 8 spectral windows of 10~MHz bandwidth, each
divided into 1024 spectral channels with an effective channel spacing
of 0.5~km\,s$^{-1}$.  Two spectral windows, one each for LCP and RCP,
were centered at the systemic LSR (local standard of rest) velocity of
the source in each of the four transitions at 6016.746, 6030.747,
6035.092, and 6049.084~MHz.  On the second day, velocities were
shifted by 20~km\,s$^{-1}$ to confirm that features present in the
data were not artifacts of the autocorrelator.  Data from the FFTS
were of lesser quality but were used to confirm that suspected
features were not artifacts of the AK90 system.

The data were reduced using the Continuum and Line Analysis
Single-dish Software (CLASS) package.  Hanning smoothing was applied
to the data to remove the effects of correlated noise generated by
strong continuum and line sources.  Polynomial baselines were removed
from the scans.

\section{Results}

\begin{deluxetable}{lllrrl}
\tablecaption{Observed Sources \label{source-table}}
\tablewidth{\hsize}
\tablehead{
  \colhead{Source} &
  \colhead{RA\tablenotemark{a}} &
  \colhead{Dec\tablenotemark{a}} &
  \colhead{$V_\mathrm{LSR}$\tablenotemark{a}} &
  \colhead{$\sigma$\tablenotemark{b}} &
  \colhead{Notes\tablenotemark{c}} \\
  \colhead{} &
  \colhead{(J2000)} & 
  \colhead{(J2000)} &
  \colhead{(km\,s$^{-1}$)} &
  \colhead{(mJy)} &
  \colhead{}
}
\startdata
\multicolumn{6}{c}{Supernova Remnant Pointings} \\
\tableline \\[-0.1truein]
\object[3C 58]{3C58}             & 02 05 38.1 & $+$64 49 41 &      0 & 11 & X       \\
\object{Crab Nebula}             & 05 34 31.9 & $+$22 00 52 &      0 & 26 & X       \\
\object{IC 443} G                & 06 16 43.6 & $+$22 32 34 &   $-$5 &  4 & \nodata \\
\object{Sgr A East} SW\tablenotemark{d}& 17 45 38.0 & $-$29 01 15 & $-$130 & 31 & \nodata \\
\object{Sgr A East} NW\tablenotemark{d}& 17 45 40.2 & $-$28 59 45 &    130 & 41 & \nodata \\
\object{Sgr A East} S            & 17 45 43.0 & $-$29 01 25 &     60 & 34 & A?      \\
\object{Sgr A East} N            & 17 45 44.5 & $-$28 58 45 &     50 & 28 & X,A?    \\
\object{Sgr A East} SE           & 17 45 48.0 & $-$29 01 00 &     60 & 23 & X,A?    \\
\object{Sgr A East} NE           & 17 45 48.5 & $-$28 59 45 &     60 & 23 & A       \\
\object{Sgr D}                   & 17 48 52.7 & $-$28 11 18 &      0 & 12 & \nodata \\
\object[SNR 1.4-0.1]{G1.4$-$0.1} & 17 49 28.1 & $-$27 47 35 &      0 & 18 & \nodata \\
\object[W 28]{W28} A             & 18 00 45.2 & $-$23 17 43 &     10 &  9 & \nodata \\
\object[W 28]{W28} C             & 18 01 34.7 & $-$23 24 00 &     10 & 18 & \nodata \\
\object[W 28]{W28} F             & 18 01 51.6 & $-$23 18 58 &     10 &  9 & \nodata \\
\object[W 28]{W28} E             & 18 01 51.6 & $-$23 17 44 &     10 &  9 & \nodata \\
\object[SNR16.7+0.8]{G16.7$+$0.8}& 18 20 58.3 & $-$14 21 52 &     20 &  6 & \nodata \\
\object{Kes 69}                  & 18 33 12.0 & $-$10 00 40 &     70 &  6 & \nodata \\
\object{G27.3+0.1}               & 18 40 35.1 & $-$04 57 45 &     35 &  6 & \nodata \\
\object[3C 391]{3C391} E         & 18 49 22.6 & $-$00 57 32 &    110 &  6 & \nodata \\
\object[3C 391]{3C391} W         & 18 49 37.1 & $-$00 55 31 &    110 &  6 & \nodata \\
\object{Kes 78}                  & 18 51 24.0 & $-$00 08 23 &     85 &  6 & \nodata \\
\object[W 44]{W44} A             & 18 55 27.2 & $+$01 33 46 &     45 &  8 & \nodata \\
\object[W 44]{W44} B             & 18 56 00.9 & $+$01 12 57 &     45 &  8 & \nodata \\
\object[W 44]{W44} E             & 18 56 29.1 & $+$01 29 39 &     45 & 11 & \nodata \\
\object[W 44]{W44} D             & 18 56 29.3 & $+$01 20 29 &     45 &  8 & \nodata \\
\object[W 44]{W44} F             & 18 56 36.7 & $+$01 26 35 &     45 &  6 & \nodata \\
\object[W 51]{W51} C             & 19 22 54.1 & $+$14 15 42 &     70 &  6 & \nodata \\
\tableline \\[-0.1truein]
\multicolumn{6}{c}{Other Pointings} \\
\tableline \\[-0.1truein]
\object[W 3 (OH)]{W3(OH)}        & 02 27 03.8 & $+$61 52 26 &      0 &  8 & A,M     \\
\object{AU Gem}                  & 07 45 28.6 & $+$30 46 43 &     15 &  6 & X       \\
\object[Onsala 1]{ON 1}          & 20 10 09.1 & $+$31 31 34 &     10 &  9 & M       \\
\object{NML Cyg}                 & 20 46 25.5 & $+$40 07 00 &      0 &  9 & X       \\
\object{NGC 7027}                & 21 07 01.6 & $+$42 14 10 &     70 & 14 & X
\enddata
\tablenotetext{a}{Pointing center and center LSR velocity of
  observations.}
\tablenotetext{b}{Average rms noise in the four transitions after
  Hanning smoothing.}
\tablenotetext{c}{A: absorption, M: maser emission, X: pointing
  without 1720~MHz masers.}
\tablenotetext{d}{Includes circumnuclear disk.}
\end{deluxetable}

Table~\ref{source-table} summarizes our results.  We do not detect
masers or thermal emission in any of the 6.0~GHz transitions toward
SNRs at a typical level of 30~mJy (5\,$\sigma$).  This limit is
approximately an order of magnitude better than that previously
obtained toward southern SNRs \citep{mcdonnell07} and comparable to
the noise limits obtained in other transitions by \citet{pihlstrom07}.

The SFRs \object[W 3 (OH)]{W3(OH)} and \object[Onsala 1]{ON~1} were
observed to confirm the reliability of the system.  In both cases,
bright masers were detected at both 6030 and 6035~MHz.  Since our
spectral resolution was insufficient to resolve the maser line widths,
the reader is referred to previously-published literature for
information on the 6030 and 6035~MHz masers in these two sources.  We
note, however, that we do not see emission in W3(OH) near
$-70$~km\,s$^{-1}$ at 6030~MHz to a $4\,\sigma$ level of 30~mJy in
Stokes~I (after Hanning smoothing the spectral resolution to
1~km\,s$^{-1}$ to eliminate ringing from the bright masers), as
previously reported by \citet{fish06}.  It is not clear whether these
masers are variable or whether the original detection was an artifact
of the Effelsberg autocorrelator system.

\begin{figure}
\resizebox{\hsize}{!}{\includegraphics{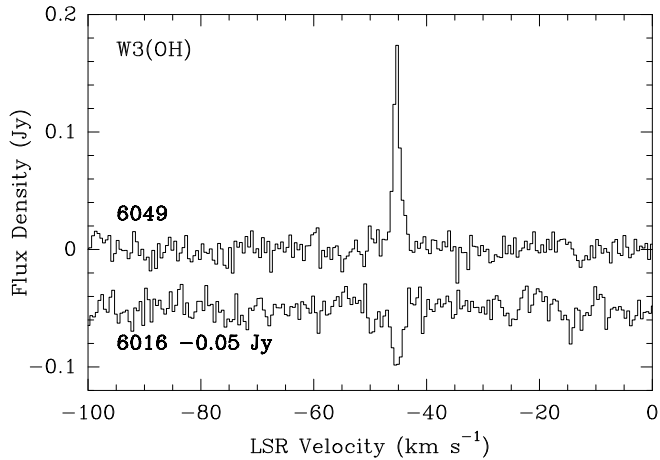}}
\caption{Spectra of the 6016 and 6049~MHz transitions in
W3(OH).  Emission in the 6049~MHz transition is stronger than
absorption in the 6016~MHz transition.  A single-Gaussian fit to the
emission is narrower than a similar fit to the absorption.
\label{fig-w3oh}
}
\end{figure}

In the satellite lines, 6016~MHz absorption and 6049~MHz emission are
seen at the same velocity toward W3(OH), as shown in
Figure~\ref{fig-w3oh}.  A single-Gaussian fit to the 6049~MHz emission
gives 165~mJy centered at $-45.30 \pm 0.03$~km\,s$^{-1}$ with an FWHM
line width of $1.39 \pm 0.08$~km\,s$^{-1}$.  These values are
consistent with those obtained by \citet{guilloteau84}, when they use
the \citet{meerts75} rest frequency of 6049.084~MHz for the
transition.  Parameters for the 6016~MHz absorption are $-52$~mJy at
$-45.22 \pm 0.13$~km\,s$^{-1}$ and a line width of $1.96 \pm
0.28$~km\,s$^{-1}$.  The emission and absorption span the same
velocity range (full width at zero power), although the peak of the
emission is narrower (FWHM) than the absorption peak, in qualitative
agreement with the observations of \citet{baudry97b}.  The agreement
in velocities suggests that the 6016~MHz absorption and 6049~MHz
emission come from the same physical region in W3(OH).

\begin{figure}
\resizebox{\hsize}{!}{\includegraphics{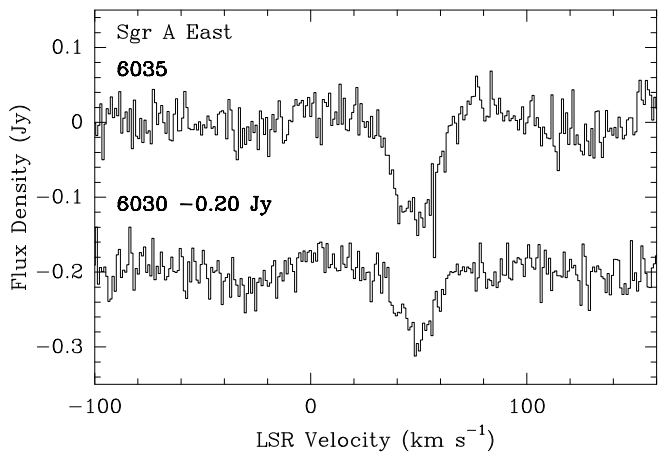}}
\caption{Spectra of the 6030 and 6035~MHz transitions in Sgr~A East.
The sharp spike near $+60$~km\,s$^{-1}$ is a correlator artifact.
Weak baseline ripples are still present after polynomial baseline
fitting.
\label{fig-ae}
}
\end{figure}

Generally no absorption is seen toward these SNRs.  Likely absorption
is detected in the main lines in the Galactic center region
(Figure~\ref{fig-ae}), although spectral baselines are poor due to
strong continuum flux.  The velocity ($\sim +50$~km\,s$^{-1}$), width
($17.5 \pm 1.5$~km\,s$^{-1}$ 6030~MHz, $20.1 \pm 1.0$~km\,s$^{-1}$
6035~MHz), and location of the strongest absorption (on the east side
of Sgr~A East) are all consistent with the densest $^{13}$CO seen in
the data of \citet{zylka96}, suggesting that excited-state molecular
content is strongest in that region.  Similar but weaker features are
seen in other scans on the eastern side of Sgr~A East, but we cannot
rule out that absorption-like features seen near $+50$~km\,s$^{-1}$ in
the other Sgr~A scans are due to errors in baseline removal.  As
discussed in \S~\ref{ae-discussion}, this absorption may be due to the
unusual molecular cloud in the region rather than its interaction with
the SNR.  Otherwise, no absorption is seen in any of the four
transitions in any of the other sources, even when scans from separate
sources (excluding the Galactic center sources) are aligned in
velocity and co-added.

We also observed \object[AU Gem]{AU~Gem} and \object[NML
Cyg]{NML~Cyg}, the only two evolved stars in which excited-state OH
emission has previously been reported \citep{zuckerman72,claussen81}.
In neither case do we detect any emission.  This is consistent with
the Expanded Very Large Array observations of \citet{sjouwerman07}
taken less than two months previously, in which no maser emission is
definitively detected in either the 6030 or 6035~MHz transitions
toward NML~Cyg.

\section{Discussion}

Though our original interest was on the 6.0~GHz OH transitions in
SNRs, we start off the discussion with our findings on our reference
source, the SFR region W3(OH).

\subsection{Satellite Lines in W3(OH)}
\label{satellite}

The 6049~MHz ($F = 3 \rightarrow 2$) emission in W3(OH) is stronger
than the 6016~MHz ($F = 2\rightarrow 3$) absorption, as noted in the
calculations by \citet{guilloteau84}, who did not detect the 6016~MHz
absorption at all.  This effect, in which 6016~MHz absorption is
enhanced relative to the expected LTE value (from 6030 and 6035~MHz
absorption) and the 6049~MHz line is seen in weak absorption or even
emission, was seen in other SFRs by \citet{whiteoak76}
and \citet{gardner83}.  This behavior was predicted by
\citet{elitzur77}, who showed that anti-inversion at 6016~MHz and
inversion at 6049~MHz should occur whenever the 120~$\mu$m radiation
linking the first-excited $J = 5/2$ state with the ground $J = 3/2$
state in the $^2\Pi_{3/2}$ ladder is optically thick, regardless of
the details of the pump mechanism.

Interestingly, a similar phenomenon is seen in the next highest state
in the $^2\Pi_{3/2}$ ladder.  Absorption in the 13442~MHz ($F = 3
\rightarrow 4$) transition is always stronger than at 13433~MHz ($F =
4 \rightarrow 3$) and always enhanced from LTE values compared to the
main lines \citep{fish05}.  The \citet{elitzur77} result is
generalizable: whenever intra-ladder decays dominate and are optically
thick, the $\Delta F = +1$ line will be seen in enhanced
absorption and the $\Delta F = -1$ line will be seen in
emission or reduced absorption.  For the 13.4~GHz transitions,
infrared overlaps in the 84~$\mu$m transitions linking the $J = 7/2$
and $5/2$ states may also be important, as proposed by
\citet{matthews86}.

We find that a Gaussian fit to the 6049~MHz emission in W3(OH) is also
\emph{narrower} than the fit to the 6016~MHz absorption, even though
the full width at zero power of the two lines is the same.  This is
consistent with weak inversion and suggests that the 6049~MHz emission
should be understood as a low-gain maser, as discussed by
\citet{baudry97b}.  Further sensitive observations at higher spectral
resolution will be required to confirm the maser nature of the
6049~MHz emission.  High angular resolution interferometric
observations would be particularly useful to determine the size and
brightness temperature of the 6049~MHz emission region, as well as to
isolate the region of emission and determine whether the inversion
exists over the entire source or merely in cluster C \citep[in the
nomenclature of][]{fish07}, which must be a region of very high
excitation based on the presence of highly-excited OH
\citep{baudry93,baudry97}.

In general, satellite-line masers are seen almost exclusively in the
lowest states of the $^2\Pi_{3/2}$ and $^2\Pi_{1/2}$ ladders.  In the
$^2\Pi_{3/2}, J = 3/2$ rotational state, masers are seen in the
1612~MHz transition in evolved stars and some SFRs, and 1720~MHz
masers are seen in SNRs and some SFRs.  Masers in the $^2\Pi_{1/2}, J
= 1/2$ state are seen only in SFRs, with 4765~MHz ($F = 1 \rightarrow
0$) masers much more common than 4660~MHz ($F = 0 \rightarrow 1$)
masers \citep[e.g.,][]{gardner83,cohen95}, in agreement with
\citet{elitzur77}.  Even though the brightest main-line 6030 and
6035~MHz masers in SFRs can have flux densities of several hundred~Jy
\citep[e.g.,][]{fish07}, the only other satellite-line maser found to
date is the low-gain 6049~MHz maser in W3(OH).  It thus appears to be
a general result that it is extremely difficult to invert satellite
lines except in the lowest rotational state in each ladder.

\subsection{Sgr~A East}
\label{ae-discussion}

We see 6030 and 6035~MHz absorption near $+50$~km\,s$^{-1}$ in
pointings on the eastern side of Sgr~A East.  The absorption at
6035~MHz is deeper than at 6030~MHz, as would be expected for thermal
absorption.  This absorption is likely neither part of the interaction
of the SNR with the molecular cloud nor located at the interaction
region \citep[with most 1720~MHz masers in the slightly higher
velocity range $V_\mathrm{LSR} =
53$--$68$~km\,s$^{-1}$;][]{pihlstrom06}, although interferometric
confirmation, as with the Expanded Very Large Array (EVLA) will be
required to establish the location of the absorption.  The molecular
material is denser to the east and partially obscured by the
circumnuclear disk to the west \citep{mcgary01,herrnstein05},
consistent with our absorption detections as well as the ground-state
absorption detected by \citet{karlsson03}.

The detection of excited-state absorption toward Sgr~A East but not
toward any other SNR may find its explanation in a chance alignment,
the special conditions in the Galactic center region, or a combination
thereof.  The difference with the other SNRs may be due to the
geometry of a dense cloud in the line of sight with a relatively
intense radio continuum background in the case of Sgr~A East,
resulting in a larger column density and absorption of OH.  Or, the
difference may be due to a difference in heating of the cloud: the
+50~km\,s$^{-1}$ cloud toward Sgr~A East may be heated more, not by
the impact of the SNR into the cloud \citep{herrnstein05} but, e.g.,
due to a strong local radiation field absorbed by high-metallicity
material, dissipation of kinetic energy or collisions of local clumps
in a steep gravitational potential, etc., providing a larger column
density of OH in excited states.  As this absorption is likely not due
to the SNR, further studies are required to investigate its origin.

Interferometric observations would be very useful in order to
understand the distribution of excited-state OH in the molecular cloud
near Sgr~A East.  Higher angular resolution than that provided by
Effelsberg will be required to understand the origin of the 6.0~GHz
absorption in Sgr~A East.  This absorption is an obvious target for
reobservation with the EVLA when a sufficient number of antennas are
equipped with the new C-band (4--8~GHz) receivers.

\subsection{Other Sources}

No emission was seen in any 6.0~GHz transition toward SNRs.  To date,
a total of 23 $\Lambda$-doublet transitions have been observed toward
SNRs: the quadruplets at 1.6, 6.0, 7.7, 8.1, and 23.8~GHz and the
triplet at 4.7~GHz.  Only the 1720~MHz transition produces a
detectable maser.  The 6049~MHz transition is the next one after
1720~MHz predicted to go into inversion as densities are increased
\citep{pihlstrom07,wardle07}, yet searches for 6049~MHz masers have
not uncovered any detections \citep[this work, as well
as][]{mcdonnell07}.  Searches for SNR OH masers at 4765 and 1612~MHz,
the next two transitions expected to produce masers via collisional
excitation, have also produced only non-detections
\citep{pihlstrom07}.  It is probable that no OH masers, other than in
the 1720~MHz transition, will be detected toward SNRs with current
instrumentation.

Absorption at 6016~MHz is not seen toward any SNR, despite predictions
that it should be anti-inverted.  We do not see absorption toward SNRs
with the exception of 6030 and 6035~MHz absorption toward Sgr~A East.
Much more sensitive observations will be required to test whether
excited-state $\Delta F = +1$ transitions are in enhanced absorption
in SNRs.

No emission was seen toward the evolved stars AU Gem and NML Cyg.
Excited-state OH maser emission has previously been reported in the
4750~MHz transition toward AU Gem \citep{claussen81} and the 6035 (and
possibly 6030) MHz transition toward NML Cyg \citep{zuckerman72},
although subsequent observations have failed to confirm these
detections \citep[e.g.,][]{sjouwerman07}.  It is likely that
excited-state OH maser emission from evolved stars is temporary and
extremely rare \citep{sjouwerman07}.  Our continued non-detection of
excited-state emission toward these stars supports this hypothesis.

\acknowledgments

Based on observations with the 100-m telescope of the MPIfR
(Max-Planck-Institut f\"{u}r Radioastronomie) at Effelsberg.  The
National Radio Astronomy Observatory is a facility of the National
Science Foundation operated under cooperative agreement by Associated
Universities, Inc.

{\it Facilities: \facility{Effelsberg}}

\end{document}